\documentclass[useAMS,usenatbib]{mn2e}
\usepackage{times,url,graphicx}

\long\def\symbolfootnote[#1]#2{\begingroup%
\def\thefootnote{\fnsymbol{footnote}}\footnote[#1]{#2}\endgroup}

\begin{document}


\title
[Keck Observations of GRO J0422+32]
{Keck Infrared Observations of GRO J0422+32 in
    Quiescence}
\author
[M.~Reynolds et al.]
{Mark T. Reynolds$^{1}$\thanks{email : m.reynolds@ucc.ie},
Paul J. Callanan$^{1}$, and Alexei V. Filippenko$^{2}$\\
$^{1}$Physics Department, University College Cork, Ireland\\
$^{2}$Department of Astronomy, University of California, Berkeley,
  California 94720-3411, USA}

\maketitle

\begin{abstract}

We present Keck K-band photometry and low-resolution H\&K-band
spectroscopy of the X-ray nova GRO J0422+32 obtained while the system
was in the quiescent state. No clear ellipsoidal modulation is present
in the light curve, which is instead dominated by a strong flickering
component. In the K-band we observe strong Br${\gamma}$
emission, with an equivalent width of 38 $\pm$ 5 {\AA}. From this we
conclude that the accretion disc is the most likely source of the
observed photometric contamination, and that previous infrared-based
attempts to constrain the mass of the putative black hole in this
system are prone to considerable uncertainty. We finally proceed to
show how it is possible to place meaningful constraints on some of the
binary parameters of this system, even in the presence of a relatively
high level of contamination from the disc. 

\end{abstract}



\section{Introduction}

Low-mass X-ray binaries (LMXBs) are systems in which a degenerate
primary, typically a neutron star or black hole, accretes from a
low-mass (M$_2$ $ < $ 1 M$_{\sun}$), late-type companion (secondary) star.
X-ray novae (XRNe) form a subset of the LMXBs: these are transient
systems which undergo periods of greatly enhanced emission, by
factors of as much as $\sim 10^6$ or greater, before returning to
quiescence over a timescale of months. However, for many of these
systems, as they return to quiescence, the secondary (typically of K
or M type) begins to make a significant contribution to the optical
flux of the system. Observations of the secondary in this state are
the most reliable means for determining the mass of the compact
object. Thus far, 20 systems have been dynamically confirmed as having
a probable black hole primary; of these, 3 are persistent sources
while the other 17 are XRNe \citep{b37}; see \citet{b5} for a thorough
review of XRNe and \citet{b2} for a more recent review of black hole binaries.

GRO J0422+32 was first detected by the \textit{Compton Gamma
Ray Observatory} (GRO) in August 1992 by \citet{b3}. The optical
counterpart was identified soon thereafter and found to have
V $\approx$ 13 mag \citep{b4}. The retreat to quiescence was
punctuated by 3 further mini-outbursts of approximately 4 magnitudes each
(\citealp{b6,b7,b8}). By September 1994, the system was deemed to have
returned to quiescence \citep{b10}, with V $\approx$ 22.4 mag, over 2 years
after the initial outburst. In total the luminosity increased by over
9 magnitudes in V, making GRO J0422+32 the highest-amplitude XRN observed
to date. \citet{b4} searched archival plates and found no
indication of any previous outburst from this system since at least 1923.

Previous attempts to determine the mass of the compact object in this
system have been made by \citet{b18}, \citet{b19}, and \citet{b12};
through optical observations, they determined the mass of the primary
to be M$_x$ = 3.57 $\pm$ 0.34~M$_{\sun}$, 2.5$M_{\sun} < $M$_x < 5 M_{\sun}$,
and M$_x > 2.2 M_{\sun}$, respectively. Recent work in the infrared (IR)
by \citet{b13} has led to a value of M$_x$ = 3.97 $\pm$ 0.95~ M$_{\sun}$;
however, negligible contamination arising from the accretion disc was
assumed. The observations we describe here were undertaken in
an effort to constrain the IR contamination due to the accretion disc,
and hence to determine the reliability of the current mass estimate
for this black hole candidate.


\section{Data}

IR photometry and spectroscopy were undertaken using the Near Infrared
Camera (NIRC; \citealt{b1}) at the f/25 forward Cassegrain port of
the Keck-I 10-m telescope. NIRC is a 256 $\times$ 256 pixel InSb array
with an angular scale of 0.15${\arcsec}$ pixel$^{-1}$. Low-resolution
spectra were obtained using a grism (R $\approx$ 120) and the HK
filter, which covers the standard H and K-bands.


\subsection{Photometry}

Photometry of GRO J0422+32 was undertaken on the night of 1997
November 9 UT. The system was observed in the K-band for a
full orbital cycle of $\sim$5.1 hr. Conditions were excellent
throughout the night, with seeing no worse than 0.45${\arcsec}$ at
any time. Individual exposure times were 4~s, with 30 coadds per
image (i.e., 2~min exposure per image). The telescope pointing was
dithered (using a 3 $\times$ 3 grid) to allow for accurate
subtraction of the IR background; see Table \ref{obs-table} for
a log of observations. A single 9-image grid of observations of
similar duration was also taken nearly a year later (1998 September
28 UT), contemporaneously with the spectroscopy (see below).

\begin{figure}
\begin{center}
\includegraphics[height=72mm,width=84mm]{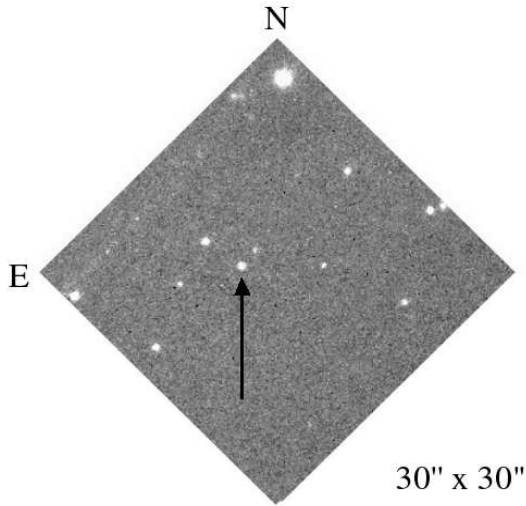}
\caption{The NIRC GRO J0422+32 field. The black arrow indicates
the position of GRO J0422+32. The exposure time was 2 min.}
\label{chart}
\end{center}
\end{figure}

\subsubsection{Data Reduction}

The data were dark-current subtracted, flat fielded, background
subtracted, and combined using standard IRAF routines\footnote{IRAF
is distributed by the National Optical Astronomy Observatories,
which are operated by the Association of Universities for Research
in Astronomy, Inc., under cooperative agreement with the National
Science Foundation}. Photometry of GRO J0422+32 and a number of
nearby stars was carried out using the profile-fitting task DAOPHOT
\citep{b11}. The standard stars FS12 and
FS16\footnote{www2.keck.hawaii.edu/inst/nirc/UKIRTstds.html}
\citep{b39} were also
observed to allow accurate calibration of the target frames. The
magnitudes were cross-checked using the bright star to the north
(2MASS J042142.57+325449.5), which is visible in the
2MASS\footnote{The Two Micron All Sky Survey is a joint project
of the University of Massachusetts and the Infrared Processing
and Analysis Center, California Institute of Technology, funded
by NASA and the National Science Foundation.} field of GRO J0422+32.
We display one of our frames in Fig. \ref{chart}. The resulting
light curve was then phased to the ephemeris of \citet{b12}. As our
data was taken $\sim$ 125 days after the ephemeris of \citet{b12}, any
errors will be minimal.

\begin{figure}
\begin{center} 
\includegraphics[height=84mm,width=84mm,angle=-90]{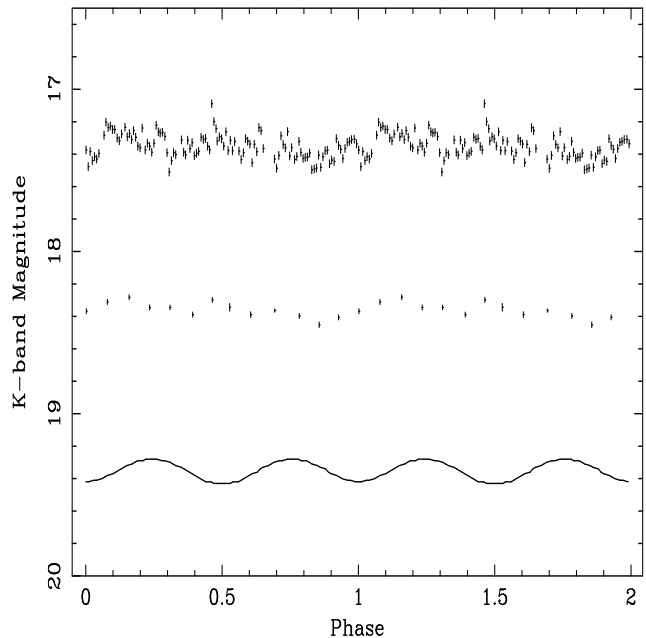}
\caption{The NIRC K-band light curve of GRO J0422+32. The top
light curve is the 2-min resolution K-band light curve. We
plot this binned to a resolution of 18~min in the middle. As a
comparison, the best-fit model of \citet{b13} is plotted below.
The unbinned light curve is accurately calibrated, while the
middle and bottom light curves are shifted downward by 1 and 2 mag
(respectively) for clarity.}
\label{alllc}
\end{center}
\end{figure}

We find that our magnitudes are in agreement with those of
\citet{b13}, confirming that our observations were indeed taken in
quiescence. It is apparent from our images that there is a
line-of-sight star to the northwest, lying approximately 2.6$\arcsec$ from
GRO J0422+32, which may have been unresolved in previous observations of
the system (see Fig. \ref{chart}). Profile fitting was carried out on
this star simultaneously with our target star, GRO J0422+32. We find
this star to have a mean magnitude of 19.62 $\pm$ 0.06, which was
constant over the course of our observations.

The photometric accuracy of our GRO J0422+32 light curve (2min
resolution) was determined
by applying a linear fit to the light curve of the star to the northeast
(Fig. \ref{chart}), of comparable magnitude to GRO J0422+32, which was
assumed to be constant during the course of our observations. The
uncertainty was then taken to be the root-mean-square (rms) deviation
from this linear fit and was measured to be 0.024 magnitudes.

The 1998 observation was reduced in a manner identical
to that described above. When phased and calibrated similarly to the
previous epochs data, the magnitude was found to be consistent. This
confirms that the system was in a quiescent state similar to that
detected the previous year.

\begin{table*}
\caption{NIRC Observation Log.}
\label{obs-table}
\begin{center}
\begin{tabular}{ccccc}
\hline
JD$^a$ & Seeing [$\arcsec$] & Exp. time [s] & $\phi^b$ & Apparent
K-band mag$^c$ \\
\hline\hline
2450762.91995413 & 0.40$\arcsec$ & 1080 &  0.528 & 17.344 $\pm$ 0.014\\
2450762.93619041 & 0.35$\arcsec$ & 1080 &  0.605 & 17.389 $\pm$ 0.010\\
2450762.95500179 & 0.35$\arcsec$ & 1080 &  0.693 & 17.364 $\pm$ 0.015\\
2450762.97392475 & 0.30$\arcsec$ & 1080 &  0.783 & 17.397 $\pm$ 0.014\\
2450762.98945141 & 0.30$\arcsec$ & 1080 &  0.856 & 17.452 $\pm$ 0.015\\
2450762.00493642 & 0.30$\arcsec$ & 1080 &  0.929 & 17.406 $\pm$ 0.015\\
2450763.02045906 & 0.30$\arcsec$ & 1080 &  0.002 & 17.368 $\pm$ 0.015\\
2450763.03688440 & 0.30$\arcsec$ & 1080 &  0.079 & 17.311 $\pm$ 0.015\\
2450763.05382848 & 0.30$\arcsec$ & 1080 &  0.159 & 17.282 $\pm$ 0.015\\
2450763.06980607 & 0.35$\arcsec$ & 1080 &  0.234 & 17.345 $\pm$ 0.014\\
2450763.08537309 & 0.40$\arcsec$ & 1080 &  0.309 & 17.345 $\pm$ 0.015\\
2450763.10399760 & 0.40$\arcsec$ & 1080 &  0.392 & 17.389 $\pm$ 0.015\\
2450763.11857811 & 0.45$\arcsec$ & 1080 &  0.464 & 17.298 $\pm$ 0.023\\
2451085.99342259 & 0.35$\arcsec$ & 1080 &  0.310 & 17.264 $\pm$ 0.093\\
\hline
\end{tabular}
\end{center}
\medskip
{$^{a}$Exposure time consists of nine 2-min exposures; JD is at the
  midpoint. \\
$^{b} T_0$ = 2450274.4156, $P$ = 0.2121600 days \\
$^{c}$Magnitudes correspond to those of the binned light curve; see
  Fig. \ref{alllc}.}
\end{table*}


\subsection{Spectroscopy}
We obtained a number of low-resolution NIRC spectra of
GRO J0422+32 on the night of 1998 September 28 UT. The HK
filter was used (1.40--2.53~$\mu$m) in conjunction with
the gr120 grism and a 0.67$\arcsec$ slit, giving a resolution
R $\approx$ 120. Observing conditions were once again excellent,
with seeing as good as 0.30$\arcsec$, for the majority of the
observations. The system was observed from orbital phase 0.36 to
0.78. Individual exposure times were 200~s, dithered to 5
positions along the slit. In total 30 spectra of the system
were acquired.

\subsubsection{Data reduction}
The one-dimensional spectra were extracted in the standard manner
using IRAF. Of the 30 exposures, 28 contained useful spectra.
Spectra of the G0V stars BS8455 and BS1789 were also taken to
aid with the removal of telluric features. No arc spectra were
taken due to the availability of a known calibration, which was
obtained from the NIRC online
manual\footnote{http://www2.keck.hawaii.edu/inst/nirc/}.

The GRO J0422+32 and G0V star spectra were normalised using a
low-order spline. The region surrounding the Br$\gamma$ feature
in the G0V spectrum was masked and the GRO J0422+32 spectra were
then divided by the G0V spectra to remove the telluric features.
The resulting spectra were median combined so as to negate the effect
of cosmic-ray hits and other spurious effects, while at the same time
maximising the available signal-to-noise ratio (S/N).

\begin{figure}
\includegraphics[height=84mm,width=64mm,angle=-90]{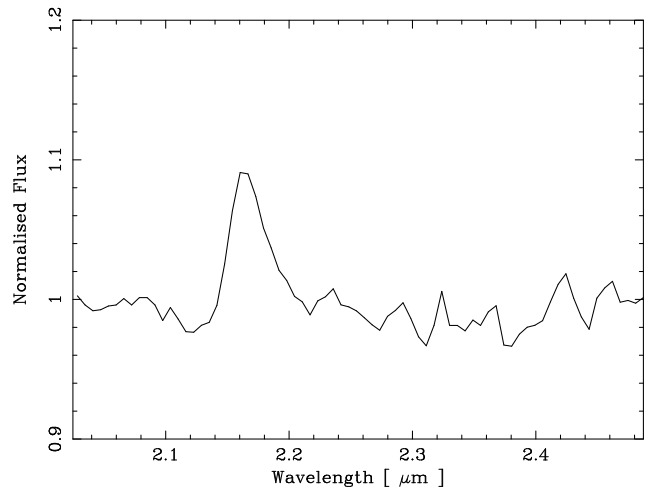}
\caption{The NIRC $K$-band spectrum of GRO J0422+32.}
\label{kspec}
\end{figure}

No significant features were detected in the resulting H-band
spectrum. In the K-band we observe strong Br${\gamma}$ emission,
which we display in Fig. \ref{kspec}. The EW of the Br${\gamma}$
feature was measured to be 38 $\pm$ 5 {\AA}.

\section{Analysis}
\subsection{Colour}
We calculated the H-K colour index of the system from our
spectra. This was achieved by fitting a polynomial to the continuum of
the individual H\&K-band spectra and then integrating the flux across
the respective passband. The median colour was found to be H-K  -0.02
$\pm$ 0.01 mag, where the error is the rms deviation of the
data. Correcting for interstellar extinction using a value of E(B-V) =
0.3 $\pm$ 0.1 mag (\citealt{b36}) for the colour excess leads to an
extinction-corrected colour of (H-K)$_0$ = 0.07 $\pm$ 0.08 mag,
which equates to spectral types in the range $\sim$ B5 to $\sim$ K7
\citep{b20}. This is somewhat earlier than the expected spectral type
of $\sim$ M2 estimated by (\citealt{b18}), although the uncertainties
are large.

\subsection{Power Density Spectrum}
It is immediately apparent in the 2-minute resolution light curve
(Fig. \ref{alllc}) that  any ellipsoidal modulation in the data is
masked by a strong flickering component. In an attempt to decouple the
ellipsoidal variation of the secondary star from the flickering
component, we followed the method of \citet{b23}. As the data sampling
was not uniform a Lomb-Scargle periodogram \citep{b24} of the 2 minute
resolution data was calculated. A significant signal ($\sim$ 5$\sigma$) is
detected at a period $\sim$ 2.5 hrs, $\sim$ half the orbital period.

The amplitude of the modulation at this period is only 0.03
magnitudes. However, as can be clearly observed from the binned (18
minute resolution data: Fig.\ref{alllc}), this variation cannot be
ellipsoidal in origin: the minima for this modulation occur at phases
$\sim$ 0.3 \& 0.8 and not at phases 0.0 \& 0.5 as would be expected
for an ellipsoidal modulation. This phase shift is much greater than
the error in calculating the orbital phase from the known ephemeris
of \citet{b12}. Hence, we can use our data to place a conservative
upper limit of 0.03 magnitudes on the amplitude of any ellipsoidal
variation present in our data.

\begin{figure}
\includegraphics[height=84mm,width=64mm,angle=-90]{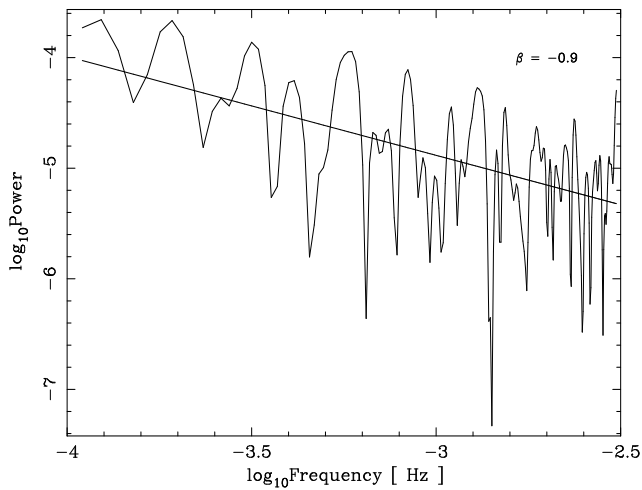}
\caption{The IR power spectrum of GRO J0422+32. The solid line
  indicates the best power law fit to the data.}
\label{logpsd}
\end{figure}

In an attempt to characterise the shorter timescale variability
present in our light curve, we computed the Power Density Spectrum
(PDS). A power law fit to our data of the form $f(x) \propto x^\beta$
yielded a power law index $\beta \approx$ -0.9: see Fig. \ref{logpsd}.
This is consistent with the variability timescales observed in the
optical in both A0620-00 and GRO J0422+32 by \citet{b23}, who measured
significant variability on timescales of up to 90 minutes.

\section{Discussion}
In the JHK photometry of \citet{b13}, an ellipsoidal modulation is
present in both sets of K-band observations (from 2000 and
2003). By contrast, our unbinned light curve (resolution $\sim$2 min),
while showing rapid variability, appears to display little evidence
of an ellipsoidal modulation. Binning the data to
$\sim$18 min resolution (this is equal to the duration of our 3
$\times$ 3 dither pattern and comparable to that of \citealt{b13}),
gives rise to a modulation that, although superficially ellipsoidal in
nature, clearly is not (see section 3.2). \citet{b14} observed this
system in 2002 with the 3.5m WIYN telescope, they also report a
modulation on a $\sim$ 5 hour period that is not ellipsoidal in nature.
It is unclear why our results differ from those of \citet{b13},
especially as the mean K-band magnitude we measure is consistent with
theirs (and hence there cannot have been a significant change in the
overall contribution of the flux from the disc to the light curve) but
it may be that their K-band modulation is similar to the one we
observe, but coincidentally at a phase that made it appear ellipsoidal
in origin.

In any case we now discuss the likely sources of the lightcurve we observe:
$(1)$ spots on the M-type secondary star, $(2)$ contamination originating in
a radio jet and/or  $(3)$ contamination from the quiescent accretion disc.

\subsection{Spots on the M-type Secondary}
Deviations from a pure ellipsoidal modulation have sometimes been
attributed to the presence of spots on the rapidly rotating
secondary star \citep{b61,b58,b57}.
We searched the literature in an attempt to find examples of M-type
dwarfs containing spots, but the number of candidate systems is few,
given that we require a rapidly rotating companion: $v_{rot} \sin(i)$
for GRO J0422+32 is close to 90 km s$^{-1}$ \citep{b50}.
The two best candidates are the isolated M1 stars HK Aqr and RE
1816+541, with rotational velocities of $\sim$ 70 km s$^{-1}$ and
$\sim$ 60 km s$^{-1}$, respectively.

Detailed V-band light curves of HK Aqr \citep{b27} show a $\sim$ 0.09
mag modulation due to spots. However, this modulation appears to be quite
uniform and smooth, in contrast to the rapid variability which we
observe in GRO J0422+32. Indeed, recent Doppler images \citep{b28}
reveal the presence of a large number of star spots on the surface
of these rapidly rotating M-type stars even though the observed variability
is quite low. Hence, we do not believe that star spots
offer a satisfactory explanation for the observed rapid variability of
GRO J0422+32 in the IR.

\subsection{Contamination from a Jet}
In recent years evidence has emerged that the synchrotron flux from a
radio jet could be contributing to the flux emitted at other
wavelengths, in particular the IR and optical \citep{b41}. IR
contamination, which has been attributed to a jet, has been observed
in the halo black hole candidate XTE J1118+480 during outburst
\citep{b53}. \citet{b43} claim that all black hole XRBs will become
jet dominated in the quiescent state. To date emission from a radio
jet in quiescence has been detected for two XRBs, V404 Cyg (GS
2023+338, \citealt{b54}) and A0620-00 \citep{b55}. The observation of
A0620 is the more pertinent here due to the similarity of its orbital
period (7.8 hrs) and quiescent X-ray luminosity ($3 \times 10^{30}$
erg s$^{-1}$) compared to those of GRO J0422+32 (5.1hrs and $8 \times
10^{30}$ erg s$^{-1}$ respectively). However, recent \textit{Spitzer}
observations \citep{b56} do not appear to be consistent with emission
from the jet in A0620-00 longwards of 8 microns: shortwards of 8
microns \citet{b56} argue that the emission appears to be consistent
with that originating from an optically thick blackbody and not the
photosphere of the secondary star. Hence, by analogy with A0620-00, we
think it unlikely that the flux we observe in the case of GRO J0422+32
originates in a jet.

\subsection{Contamination from the Accretion Disc}

\subsubsection{Photometric Constraints}
Previous estimates of the fractional contribution of the accretion
disc to the total flux range from $\sim$60\% \citep{b18} to as little
as 20\% \citep{b33} in the R-band. In a subsequent re-analysis of the
data from \citet{b18}, \citet{b50} measured the flux from the
accretion disc to contribute $\sim$ 40\% of the total optical flux in
quiescence. Similarly \citet{b12} measured the contamination from the
accretion disc to be $\sim$ 60\% in the I-band. These measurements
hint at an increasing level of contamination at longer wavelengths,
although the different template stars used in the respective analyses
(M2 vs M4) limit the conclusions that can be made.

In their analysis of GRO J0422+32, \citet{b13} assumed a negligible
contribution from the accretion disc in the K-band.
To check this assumption, we have used the ELC light curve modelling
code \citep{b21} to model the flux from an accretion
disc, assuming a temperature profile of the form $T(r) \propto
T_{inner}(r/r_{inner})^{\xi}$, with $\xi$ chosen so as to ensure a
steady state disc, and an outer disc radius of $\sim$ 0.55 R$_L$
(these values being consistent with observations of the quiescent disc
in the XRNe A0620-00 and XTE J1118+480 (\citealt{b25,b59,b60}).
We calculate that for any value of the accretion disc contribution to
the total flux in the R-band, the fractional contribution of the
accretion disc flux to the K-band is similar, if not greater.  
This is in contrast to the canonical picture where the flux from the
disc falls off as we venture towards the IR. The
contributions from the accretion disc measured in the R \& I-band in
the case of GRO J0422+32 agree with this simple model and hence we
would expect a similar level of contamination in the K-band.  This
result is dependent on the inclination of the system, as the flux from
the accretion disc relative to that from the secondary increases with
lower inclination.

\begin{figure}
\includegraphics[height=84mm,width=64mm,angle=-90]{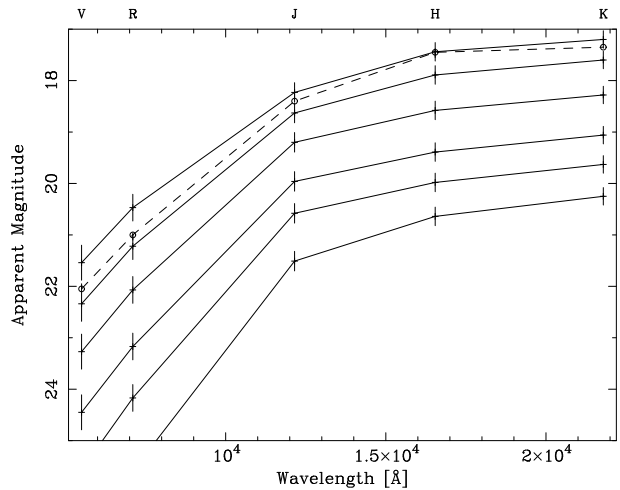}
\caption{VRJHK SED of various M-type stars and GRO J0422+32 (dashed line).}
\label{rel-flux}
\end{figure}

In Fig. \ref{rel-flux}, we plot the VRJHK spectral energy distribution
(SED) for various M-type secondary stars along with that of GRO
J0422+32. A distance of 2.6 $\pm$ 0.2 kpc and a reddening of E(B-V) =
0.3 $\pm$ 0.1 have been assumed (from \citet{b51}). The absolute
V-band magnitudes and the various colour relations of the M-type stars
are taken from \citet{b52}. The dashed line indicates the observed SED
of GRO J0422+32. The error bars account for both the uncertainty in
the reddening and the distance values quoted above. The observed
colours are broadly consistent with a spectral type of M1, and suggest
a disk contribution in the K-band of no more that $\sim$ 0.3 mag
(albeit at the 1$\sigma$ level). This value would be consistent with
that observed in the R-band, as expected from the modelling discussed
above. Alternatively, if there existed a
cool optically thick component to the accretion disc (as might be
expected theoretically - see \citealt{b63}), a contribution from a radio
jet (see Section 4.2 above), or some form of circumbinary disc
(\citealt{b56}), any of these contributions could lead to a flatter SED
than might otherwise be expected. 

A spectral type of M1 is consistent with previous spectral type
determinations (\citealt{b19,b18,b13}), with the exception of
\citet{b12} who measured a spectral type between M4 \& M5.  To
reconcile the latter spectral type with the observed SED would require
that the distance estimate used above is significantly in error, or a
K-band contamination from the accretion disk of $\sim$ 3 magnitudes
(which we also regard as unlikely).

\subsubsection{Spectroscopic Measurements}

The Br${\gamma}$ feature at 2.16$\mu$m, with a equivalent width of
$\sim$ 38 \AA, dominates the K-band spectrum of GRO J0422+32.  No
other spectral lines are visible in either the H or K-bands. We also
note the absence of the CO bandheads from the companion star in this
spectrum, although it is possible that they are simply unresolved.

Previous observations of LMXBs support the conclusion that the
presence of strong Br${\gamma}$ emission is due to contamination by
the accretion disc.  IR spectroscopy of a sample of persistently
bright neutron star LMXB \citep{b46,b15,b49} revealed the presence of
Br${\gamma}$ emission. In the sources with the strongest emission (Sco
X-1, Sco X-2, GX 5-1), the presence of a large Br${\gamma}$ feature
(EW $\sim$ 23, 29, 45 \AA) was accompanied by the absence of the CO
bandheads or any other spectral features due to the secondary
star. In contrast, in the sources with lower levels of Br${\gamma}$
emission (GX 1+4, GX 13+1, EW $\sim$ 10\AA, 5\AA), the CO bandheads
and other lines due to the secondary star were detected.

Previous quiescent IR observation of the XRNe V404
Cyg and A0620-00 \citep{b47,b48} also reveal the presence of
Br${\gamma}$ emission. In V404 Cyg the equivalent width of the
Br${\gamma}$ line was measured to be only 2.9 \AA. In this case the CO
bandheads from the secondary star were also detected. The
Br${\gamma}$ line was detected with an EW $\sim$ 15 \AA~ in
A0620-00. In this case the CO bandheads were only marginally
detected (the S/N of the spectrum was very low). Using these
spectra the authors measured the IR contamination from the accretion
disc to be $\leq$ 14\% and 27\% respectively. Recent IR spectroscopy of
A0620-00 by \citet{b40} support the measurement of excess K-band
flux in this system. 

In summary, our data, both spectroscopic and photometric, strongly
suggest a significant contribution to the K-band flux of GRO J0422+32
from the accretion disc. As previously discussed, these observations
are not the first to observe ``flickering'' in the light curve of a
quiescent XRN: \citet{b31} obtained fast photometry of the XRNe
A0620--00, Nova Mus 1991, and MM Vel in the optical and detected rapid
variability similar to that reported here. \citet{b23} obtained fast
photometry of 5 XRNe including GRO J0422+32; significant variability
was found, superposed on the ellipsoidal variations. We note that of
the 5 XRNe observed optically (V404 Cyg, A0620-00, GRO J0422+32, GS
2000+25 \& Cen X-4), the largest amplitude variability was observed in
the light curve of GRO J0422+32. \citet{b23} contend that the observed
optical flaring originates in the accretion disc and not as
chromospheric activity from the secondary. Our observations are the
first to show this kind of flickering variability in the K-band, where
such contamination had previously been thought to be minimal for
quiescent XRNe. 

\subsection{Mass of the Compact Object}
As demonstrated above, there is no clear ellipsoidal modulation
present in our K-band data. However, for a given accretion disc
contamination level in the K-band, we can still hope to constrain
the mass ratio $q$, and orbital inclination, $i$, as follows. Firstly,
we assume that the true ellipsoidal modulation has an amplitude less
than the 0.03 magnitude variation discussed in Section 3.2.

\begin{figure}
\includegraphics[height=84mm,width=64mm,angle=-90]{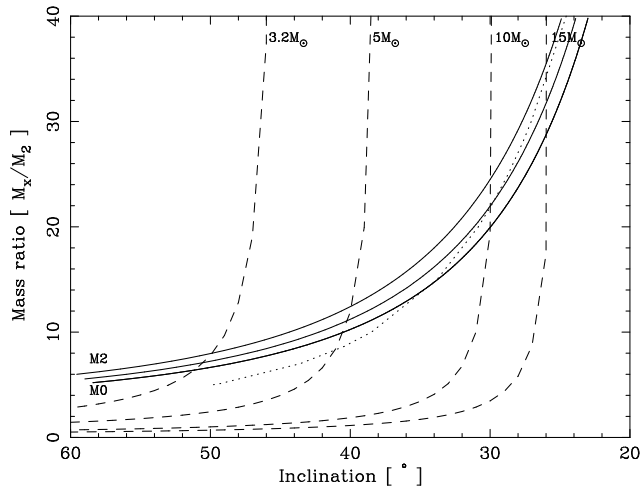}
\caption{Mass ratio inclination plot for GRO J0422+32. The dashed
  lines indicate lines of constant primary mass. The solid lines
  indicates lines of constant mass appropriate for an M0, M1 \& M2-type
  secondaries. The dotted line indicate the best fit q,i pairs for a
  0.04 amplitude sine wave. }
\label{mfunc}
\end{figure}

Assuming that the excess flux from the accretion disc contributes
approximately 30\% of the flux in the K-band (see section 4.3.1), an
intrinsic ellipsoidal modulation of 0.04 magnitudes would be reduced to
the 0.03 magnitude variation limit above. In Fig. \ref{mfunc}, we
display the $q$ vs $i$  plot for GRO J0422+32: the dotted line
indicates the best fit to a 0.04 magnitude modulation. We can
represent the ellipsoidal modulation by a single amplitude because, at
these relatively low inclinations, the depth of the minima at phases
0.0 and 0.5 are very similar. 

As this line represents an upper limit to any ellipsoidal modulation
present, allowed values of $q$ and $i$ are constrained to lie to the
right of and below this line. By combining this with the limit from
the radial velocity measurements (assuming a secondary mass of 0.45
M$_{\sun}$), as delineated by the solid line, we can constrain $q$ and
$i$ as follows: $i < 30 \degr$ and $q > 23$. Note that the inclination
constraint is an upper limit because we have only an upper limit on
the amplitude of the ellipsoidal modulation. The latter
yields a lower limit to the compact object mass of $\sim$ 10.4
M$_{\sun}$. This is in contrast to the inclination of 45$\degr \pm$
2$\degr$ (M$_x$ = 3.97 $\pm$ 0.95 M$_{\sun}$) obtained by \citet{b13},
bearing in mind the uncertainties discussed earlier.

\section{Conclusions}

The K-band properties of GRO J0422+32 are consistent with a
flux contaminated by the emission from the accretion disc, and not
solely from the secondary as previously supposed.

Previous authors have assumed negligible contamination of the IR flux
of quiescent XRNe by the accretion disc, and hence that the
ellipsoidal modulation and black hole mass could be most accurately
measured in the IR (in comparison to the optical). This conclusion was
generally based on an extrapolation of the contamination from the
optical to longer wavelengths (see \citealt{b22}). Our data show that
this is apparently an invalid assumption in the case of GRO J0422+32.

Our observations are the first to detect flickering from a quiescent
XRN accretion disc in the K-band, most likely because of the superior
S/N obtainable with the 10-m Keck telescope compared with the smaller
telescopes from which similar measurements have been previously
obtained. However. it is also possible that the K-band flickering is
present because GRO J0422+32 is a relatively low inclination system:
in this case, if the source of the flickering is located in the inner
disc, then the flickering would be more easily observable in a low
inclination system such as GRO J0422+32 in comparison to other (higher
inclination) quiescent XRNe.

In light of the observations discussed in this paper, it is clear that
higher time resolution photometry or higher S/N spectroscopic
observations of other XRNe in the IR are required to better constrain
the contribution of the accretion disc at these wavelengths, and to
determine the degree to which IR ellipsoidal variability measurements can be
used to reliably constrain the mass of black holes in quiescent
XRNe. Flickering of the amplitude reported here (0.2 magnitudes) would
have a significant effect on even the highest amplitude ellipsoidal
variation ($\sim$ 0.3 magnitudes: see for example the case of GRO
J1655-40 discussed by \citealt{b62}). Even when such contamination is
significant, however, it may still be possible to extract usefull
limits on system parameters, as long as the degree of such
contamination can be quantified (e.g. spectroscopically), as we have
discussed here.

\bigskip

This research made use of the SIMBAD database, operated at CDS,
Strasbourg, France, and NASA's Astrophysics Data System.
The data presented herein were obtained at the W. M. Keck
Observatory, which is operated as a scientific partnership
among the California Institute of Technology, the University
of California, and the National Aeronautics and Space Administration.
The Observatory was made possible by the generous financial support
of the W. M. Keck Foundation. A.V.F. is grateful for support
from NSF grant AST-0307894. M.T.R. \& P.J.C. acknowledge financial support
from Science Foundation Ireland. We thank an anonymous referee for
useful comments which improved the quality of this paper.



\bsp


\end{document}